\documentclass[fleqn,number,sort&compress,3p]{elsarticle}
\usepackage[english]{babel}
\usepackage{amsmath}
\begin{document}
\title{A manifestly gauge-invariant description of interaction of atomic systems with strong fields in the dipole approximation}
\author{A. Bechler\corref{cor1}}
\ead{adamb@univ.szczecin.pl}
\author{M. \'Sl\c{e}czka}
\address{Institute of Physics, University of Szczecin, Wielkopolska 15, 70-451 Szczecin, Poland}
\cortext[cor1]{Corresponding author}
\begin{abstract}
We propose a new type of gauge-invariant expansion of the ionization probability amplitudes of atoms by short pulses of electromagnetic radiation. Contrary to previous gauge-invariant approaches to this problem it does not require different partitions of the total Hamiltonian depending on the choice of gauge. In a natural way the atomic potential is treated as perturbation acting on an electron interacting with strong pulse. Whereas this is a standard assumption of strong field approximation (SFA), we show that grouping consequently together \textit{all} terms of the same order in the atomic potential results in the expansion of the amplitude which is gauge invariant \textit{order by order}, and not only in the limit of infinite series. In this approach, which is illustrated by numerical examples, the "direct ionization" and "rescattering" contributions are different from those commonly used in SFA - calculations.
\end{abstract}
\maketitle
\section{Introduction}
Theoretical descriptions of the interaction of atomic and molecular systems with strong external fields is based mainly on the Keldysh-Faisal-Reiss (KFR) theory~\cite{Keldysh,Faisal,Reiss} where, in the first approximation, influence of atomic forces on the electron dynamics is neglected, except of the exact initial state wave function. Assumption about dominance of the external field over atomic forces is justified in the case of very strong electric fields of the laser beam, basically of the order of the atomic electric field or higher. For a recent application of the Keldysh theory to the description of ionization by strong one-cycle pulses see, for instance,~\cite{Karnakov}. This type of approximation, known also as strong field approximation (SFA), has been applied to the description of interaction both with long optical wave-trains containing many cycles of the carrier wave~\cite{Becker1}, and with relatively short few-cycle pulses~\cite{Brabec,Milosevic}. Theoretical description in the framework of SFA including the rescattering contribution leads to a fair agreement with experimental data for above threshold ionization (ATI) spectra~\cite{Milosevic1,Gazibegovic}. The KFR theory modified to account for asymptotic Coulomb effects leads also to a qualitative agreement with the measurements of momentum distributions in the direction of linear polarization of the pulse electric field~\cite{Faisal1,Faisal2}.

One of main theoretical problems of SFA, despite of its well established status as a tool used for description of interaction with strong fields, is connected with the proper choice of gauge of electromagnetic potential describing the field of the laser pulse. Two most popular gauges, the velocity gauge (VG) and the length gauge (LG),
lead in general to different predictions~\cite{BauerJ,BauerD,Reiss1}. In some cases the length-gauge SFA-calculation for detachment from negative ions matches better results obtained by numerical solution of time-dependent Schr\"odinger equation (TDSE)~\cite{BauerD}, this gauge seems also to be "favored" by experimental results, giving predictions in agreement with measurements, at least quantitatively~\cite{Bergues,Gazibegovic1,Kjeldsen}.
This conclusion has been questioned by Reiss~\cite{Reiss1}, who argued that velocity-gauge is more suitable for the description of experimental results for detachment fron negative fluorine ion, reported in~\cite{Bergues}. Velocity gauge was earlier advocated for by Cormier and Lambropoulos in~\cite{Cormier}. The question of getting gauge invariant predictions in the case of interaction with few-cycle pulses in VG, LG and Henneberger frame~\cite{Henneberger} was discussed in the context of unitary transformations in~\cite{Madsen}.

Whereas gauge invariance of transition amplitudes (up to a phase factor) and probabilities does not raise questions as such, and can be easily proved formally, this fundamental property may be violated by various types of approximate calculations.
In the theoretical description of the interaction with strong laser fields two approximations are used: dipole and SFA, and both can be considered as sources of gauge-dependence. Obviously, the question of "$\mathbf{p}\cdot\mathbf{A}$" vs. "$\mathbf{d}\cdot\mathbf{E}$" (VG vs. LG) description of  interaction with electromagnetic field in dipole approximation is not new, especially in the context of bound-bound transitions in the perturbation theory (for reference, cf. eg.~\cite{Lamb,Starace,Fried,Bassani,Kobe,Kobe1,Rzazewski}).
In the case of bound-free transitions in strong laser fields it has been shown in~\cite{Faisal3} that VG - and LG - transition amplitudes are equivalent in all orders of the expansion provided appropriate initial and final state partitions of the Hamiltonian are chosen. In general, however, expansions of the amplitude in different gauges require various gauge dependent partitions of the Hamiltonian, and lead to the same result only in the limit of infinite series~\cite{Vanne}, and not "order by order" in the expansion.\\
\indent In this paper we propose an expansion of the amplitude which does not require any special, gauge dependent partitions of the Hamiltonian, and is manifestly gauge invariant in every order separately. The main idea is to collect together consequently all terms of a given order in atomic potential treated as perurbation in comparison with strong external field. The first gauge dependent term in standard SFA expansion is, in fact, of the first order in atomic potential, as can be seen for instance in~\cite{BauerD,BeckerW}. We show in particular that next gauge dependent term in the expansion, corresponding to first order rescattering process~\cite{Milosevic}, when combined with the first one, gives a complete gauge independent contribution of the first order in atomic potential, plus a term which is of next order in atomic potential. The iterative procedure leads then to an expansion which is manifestly gauge independent order by order.
\section{The manifestly gauge inavariant expansion of the amplitude}
The Hamiltonian describing interaction of an atomic system with external strong field in single active electron approximation has the following form (atomic units are used throughout)
\begin{equation}\label{eq:1}
    \hat{H}(t)=\hat{H}_{at}+\hat{F}(t),
\end{equation}
where
\begin{equation}\label{eq:2}
    \hat{H}_{at}=\frac{\hat{\mathbf{p}}^2}{2}+\hat{V}_{at},
\end{equation}
with $\hat{V}_{at}$ denoting time-independent atomic potential, $\hat{F}(t)$ describes the interaction with external field and $\mathbf{p}$ is the canonical momentum $\mathbf{p}=-i\nabla$. The Hamiltonian \eqref{eq:1} can be equivalently written as
\begin{equation}\label{eq:3}
    \hat{H}(t)=\hat{H}_F(t)+\hat{V}_{at},
\end{equation}
where $\hat{H}_F(t)$ is the Hamiltonian of an electron in the external field, ie.
\begin{equation}\label{eq:4}
    \hat{H}_F(t)=\frac{\hat{\mathbf{p}}^2}{2}+\hat{F}(t).
\end{equation}
Using partitions of the Hamiltonian defined by \eqref{eq:1} and \eqref{eq:3} one obtains two forms of the integral equation for the time evolution operator
\begin{subequations}\label{eq:5}
  \begin{equation}\label{eq:5a}
    \hat{U}(t,t')=\hat{U}_{at}(t,t')-i\int_{t'}^t dt_1\hat{U}(t,t_1)\hat{F}(t_1)\hat{U}_{at}(t_1,t'),
  \end{equation}
  \begin{equation}\label{eq:5b}
    \hat{U}(t,t')=\hat{U}_F(t,t')-i\int_{t'}^t dt_1\hat{U}(t,t_1)\hat{V}_{at}\hat{U}_F(t_1,t'),
  \end{equation}
\end{subequations}
where $\hat{U}_{at}$ is generated by the atomic Hamiltonian \eqref{eq:2} and $\hat{U}_F$ - by the Hamiltonian \eqref{eq:4}.
Transition amplitude from an initial bound electron state $|\phi_i(t')\rangle$ to a final continuum state $|\phi_f(t)\rangle$, orthogonal to the initial state (both are eigenstates of the atomic Hamiltonian), reads
\begin{eqnarray}\label{eq:6}
    \nonumber M=\langle\phi_f(t)|\hat{U}(t,t')|\phi_i(t')\rangle\\
    =-i\int_{t'}^t dt_1\langle\phi_f(t)|\hat{U}(t,t_1)\hat{F}(t_1)|\phi_i(t_1)\rangle,
\end{eqnarray}
where $|\phi_i(t_1)\rangle=\hat{U}_{at}(t_1,t')|\phi_i(t')\rangle$. This exact expression for the amplitude is, up to a phase factor, independent on the choice of gauge. For short pulses the interaction Hamiltonian $\hat{F}(t)$ is practically zero for times earlier than some initial time $t_i$ and later than a final time $t_f$. Therefore, the time integration limits in \eqref{eq:6} can be replaced by $t_i$ and $t_f$.

Expansion of the amplitude in powers of atomic potential can be achieved by consecutive iterations of \eqref{eq:5b}. First term corresponds to the replacement of the evolution operator by $\hat{U}_F$, so that
\begin{equation}\label{eq:7}
    M\approx M_0=-i\int_{t_i}^{t_f} dt_1\langle\phi_f^a(t)|\hat{U}_F(t,t_1)\hat{F}(t_1)|\phi_i(t_1)\rangle,
\end{equation}
where also an approximation $|\phi_f^a(t)\rangle$ for the final state has been used. Final state can be approximated by plane wave~\cite{Reiss,Milosevic},which is well justified for ionization of negative ions. In the case of ionization of a neutral atom one should use rather a plane wave distorted by asymptotic Coulomb phase~\cite{Faisal1,Faisal2}. The initial and approximated final state are not, in general, orthogonal to each other. Using equation fulfilled by $\hat{U}_F$,
\begin{equation}\label{eq:8}
    -i\partial_{t_1}\hat{U}_F(t,t_1)=\hat{U}_F(t,t_1)\left[\frac{\hat{\mathbf{p}}^2}{2}+\hat{F}(t_1)\right],
\end{equation}
substituting the product $\hat{U}_F\hat{F}$ from \eqref{eq:7} into \eqref{eq:8}, performing integration by parts and using the equation fulfilled by $|\phi_i(t_1)\rangle$, one finds
\begin{eqnarray}\label{eq:9}
    \nonumber M_0=\langle\phi^a_f(t_f)|\hat{U}_F(t_f,t_i)|\phi_i(t_i)\rangle-\langle\phi^a_f(t_f)|\phi_i(t_f)\rangle\\
    -i\int_{t_i}^{t_f} dt_1\langle\phi_f^a(t_f)|\hat{U}_F(t_f,t_1)\hat{V}_{at}|\phi_i(t_1)\rangle.
\end{eqnarray}
In the case of a periodic wave train or long pulses the boundary term gives a negligible contribution and the amplitude is then determined solely by the third term~\cite{Reiss,BauerD,BeckerW}.

Gauge transformation is implemented by a unitary operator of the form $\exp[i\chi_g(\mathbf{r},t)]$, where for transformations compatible with the dipole approximation $\chi_g$ can be at most linear in the coordinate $\mathbf{r}$. Vector potential, scalar potential and the evolution operator transform according to
\begin{eqnarray}\label{eq:10}
    \nonumber\mathbf{A}_g=\mathbf{A}-\nabla\chi_g,\quad\varphi_g=\varphi+\partial_t\chi_g,\\
    \hat{U}_g(t,t')=e^{i\chi_g(t)}\hat{U}(t,t')e^{-i\chi_g(t')}.
\end{eqnarray}
Transformed Hamiltonian has the form
\begin{equation}\label{eq:11}
    \hat{H}_g(t)=e^{i\chi_g(\mathbf{r},t)}\hat{H}(t)e^{-i\chi_g(\mathbf{r},t)}-\partial_t\chi_g(\mathbf{r},t).
\end{equation}
It follows from \eqref{eq:11} that the interaction Hamiltonian in a new gauge reads
\begin{eqnarray}\label{eq:11a}
    \nonumber\hat{F}_g(t)=\hat{F}(t)-(1/2)(\hat{\mathbf{p}}\cdot\nabla\chi_g+\nabla\chi_g\cdot\hat{\mathbf{p}})\\
    +(1/2)(\nabla\chi_g)^2-\partial_t\chi_g.
\end{eqnarray}
The Hamiltonian in a new gauge is again partitioned either according to \eqref{eq:1} or according to \eqref{eq:3},
\begin{equation}\label{eq:11b}
    \hat{H}_g(t)=\hat{H}_{at}+\hat{F}_g(t)=\hat{H}_{F_g}(t)+\hat{V}_{at}.
\end{equation}
 We shall consider a class of gauge transformations for which $\hat{\chi}_g(t)=0$ for $t\leq t_i$ and $t\geq t_f$. This class of gauge transformations is characterized by
\begin{equation}\label{eq:12}
    \hat{\chi}_g(\mathbf{r},t)=\gamma\hat{\mathbf{r}}\cdot~\mathbf{A}(t),
\end{equation}
where $\gamma$ is a real parameter\footnote{This is a subclass of a wider class of transformations considered in~\cite{Vanne}.}. Whereas first two terms in \eqref{eq:9} are gauge-independent, the third one is not since $\hat{U}_F(t_f,t_1)\neq\hat{U}_{gF}(t_f,t_1)$ for transient times.
Since the derivation leading from \eqref{eq:7} to \eqref{eq:9} can be done in any gauge, transition amplitude in the new gauge $g$ can also be written as
\begin{eqnarray}\label{eq:13}
 \nonumber M_{g0}=\langle\phi^a_f(t_f)|\hat{U}_{gF}(t_f,t_i)|\phi_i(t_i)\rangle-\langle\phi^a_f(t_f)|\phi_i(t_f)\rangle\\
    -i\int_{t_i}^{t_f} dt_1\langle\phi_f^a(t)|\hat{U}_{gF}(t_f,t_1)\hat{V}_{at}|\phi_i(t_1)\rangle.
\end{eqnarray}
First two terms in \eqref{eq:9} and \eqref{eq:13} are equal but remaining terms differ among themselves, since for transient times $\hat{U}_F(t_f,t_1)\neq\hat{U}_{gF}(t_f,t_1)$.
Next term, $M_1$, in the expansion of the amplitude can be obtained from \eqref{eq:5a} after substituting for $\hat{U}$ first iteration of \eqref{eq:5b}, which gives
\begin{eqnarray}\label{eq:14}
     \nonumber M_1=(-i)^2\int_{t_i}^{t_f}dt_1\int_{t_1}^{t_f}dt_2\langle\phi_f^a(t_f)|\hat{U}_F(t,t_2)\hat{V}_{at}\\
     \times\hat{U}_F(t_2,t_1)
     \hat{F}(t_1)|\phi_i(t_1)\rangle.
\end{eqnarray}
Changing the order of integration, using again \eqref{eq:8} to express the product $\hat{U}_F\hat{F}$ and performing integration by parts we obtain
\begin{eqnarray}\label{eq:15}
    \nonumber M_1=i\int_{t_i}^{t_f}dt_2\langle\phi_f^a(t_f)|\hat{U}_F(t_f,t_2)\hat{V}_{at}|\phi_i(t_2)\rangle\\
    -i\int_{t_i}^{t_f}dt_2\langle\phi_f^a(t_f)|\hat{U}_F(t_f,t_2)\hat{V}_{at}\hat{U}_F(t_2,t_i)|\phi_i(t_i)\rangle+...,
\end{eqnarray}
where omitted term is of the second order in atomic potential. Adding \eqref{eq:9} and \eqref{eq:15} cancels the gauge dependent contributions in both terms, leading to a manifestly gauge independent approximate expression for the amplitude of the form $M\approx M^{(0)}+M^{(1)}$, where
\begin{subequations}\label{eq:17}
    \begin{equation}\label{eq:17a}
        M^{(0)}=\langle\phi^a_f(t_f)|\hat{U}_F(t_f,t_i)|\phi_i(t_i)\rangle-\langle\phi^a_f(t_f)|\phi_i(t_f)\rangle,
    \end{equation}
    \begin{equation}\label{eq:17b}
        M^{(1)}=-i\int_{t_i}^{t_f}dt_2\langle\phi_f^a(t_f)|\hat{U}_F(t_f,t_2)\hat{V}_{at}\hat{U}_F(t_2,t_i)|\phi_i(t_i)\rangle.
    \end{equation}
\end{subequations}

\section{Numerical calculations}

We have performed numerical calculations for a strong subfemtosecond laser pulse linearly polarized along the $z$ axis. Vector potential in dipole approximation is
\begin{equation}
\mathbf{A}(t)=\frac{E_0}{\omega}\mathbf{e}f(t)\sin(\omega t+\varphi),
\end{equation}
and $\mathbf{E}(t)=-d/dt\mathbf{A}(t)$, $E_{0}$ is the electric field amplitude, $\mathbf{e}$ is the polarization direction, $\omega$ is the carrier wave frequency, $\varphi$ is the carrier-envelope phase (CEP) and $f(t)=\exp\left[ -t^{2}/2\tau^{2}\right]$ is the envelope function. The value of $\tau$ is 1.94 a.u., corresponding to FWHM of the intensity equal to 3.23 a.u. The carrier wave length used in the calculations has been chosen as 72 nm (carrier wave period $T=2\pi/\omega=240$ as) and maximum electric field as $E_0=10\ \textrm{a.u.}$, i.e. the maximum intensity $I_{max}\approx7\times10^{18}\ \textrm{W/cm}^2$. Possibilities of producing strong subfemtosecond pulses in a short wavelength region have been recently discussed eg. in~\cite{Tsakiris,Tarasevitch,Rohringer,Goulielmakis}.

We have calculated momentum distributions of final electrons for the detachment of H$^-$ ion, modeled by a short range static potential of the form~\cite{Milosevic}
\begin{equation}\label{eq:18}
V_{at}(r)=-\left(d+\frac{g}{r}\right)e^{-\mu r},
\end{equation}
with $g=d=1$ and $\mu=1.56\ \textrm{a.u.}$ This model potential supports a weakly bound state with the ionization energy $0.024\ \textrm{a.u.}$, characteristic for the H$^-$ ion. The initial bound state wave function has been found by numerical solution of the stationary Schr\"{o}dinger equation and
the amplitude was found by numerical calculation of the integrals \eqref{eq:17} in momentum space.

\begin{figure}[here]
\centering
\includegraphics[scale=0.4]{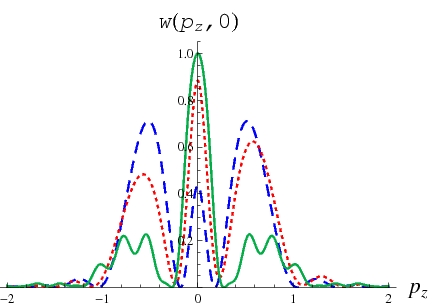}
\includegraphics[scale=0.4]{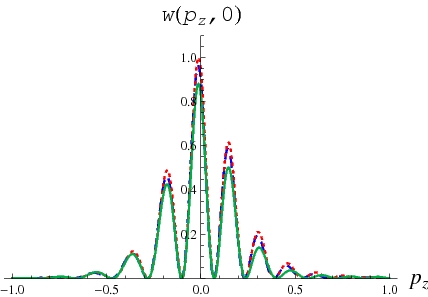}
\caption{\footnotesize{Normalized momentum distribution in the polarization direction as function of the longitudinal momentum for detachment from $\textrm{H}^-$ ion for $\varphi=0$ (left panel) and $\varphi=\pi/2$ (right panel). Blue (dashed) line - distribution calculated using $M^{(0)}$ (direct ionization), red (dotted) line - calculation with $M^{(0)}+M^{(1)}$ i.e with rescattering accounted for, green (continuous) line - standard SFA calculation in velocity gauge.}}
\label{fig_pt=0}
\end{figure}

Results of calculation of the photoelectron spectra in the direction of polarization, together with the standard SFA results in the velocity gauge for comparision, are shown in Fig. \ref{fig_pt=0}. For a pulse with carrier-envelope phase (CEP) $\varphi=0$ (left panel) the spectra calculated with the use of $M^{(0)}$, and standard SFA - approach are symmetric around $p_z=0$, whereas the correction $M^{(1)}$, to the amplitude, which includes part of a standard rescattering contribution, results in an asymmetric spectra. Spectra calculated with the use of the gauge-invariant approach (blue and red lines) differ substantially from the standard SFA results for smaller values of momentum. For $\varphi=\pi/2$ (right panel) the differences between those three approximations are much smaller.

\begin{figure}[here]
\centering
\includegraphics[scale=0.4]{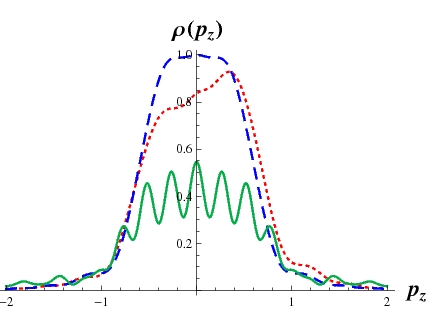}
\includegraphics[scale=0.4]{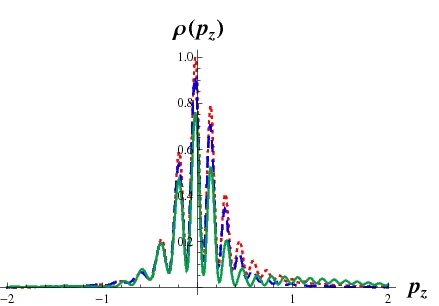}
\caption{\footnotesize{Normalized momentum distributions integrated over transverse momentum as function of longitudinal momentum for detachment from $\textrm{H}^-$ ion for $\varphi=0$ (left panel) and $\varphi=\pi/2$ (right panel). Blue (dashed) line corresponds to direct ionization, red (dotted) line takes rescattering contribution into account and green line represents conventional SFA in velocity gauge.}}
\label{fig_int_pt}
\end{figure}
Photoelectron spectra in the direction of polarization integrated over the transverse momentum, ie.
\begin{equation}\label{eq:19}
\rho(p_z)=\frac{1}{4\pi^2}\int^{\infty}_{0}w(p_z,\ p_t)p_tdp_t,
\end{equation}
are shown in Fig. \ref{fig_int_pt}.
Again, for $\varphi=0$ one can observe larger differences between gauge-invariant direct scattering, gauge-invariant rescattering and standard SFA approaches than in the case at $\varphi=\pi/2$, especially for smaller values of $p_z$. For $\varphi=\pi/2$ the spectrum shows much richer structure than for $\varphi=0$. As expected, for $\varphi=0$, ie. even electric field, $\mathbf{E}(-t)=\mathbf{E}(t)$, the "direct ionization" probablility $\propto |M^{(0)}|^2$ and the standard SFA probability are symmetric around $p_z=0$, as can be seen from left panels in Figs. \ref{fig_pt=0} and  \ref{fig_int_pt}. For $\varphi=\pi/2$ and odd electric field this property is not true.

\begin{figure}[here]
\centering
\includegraphics[scale=0.4]{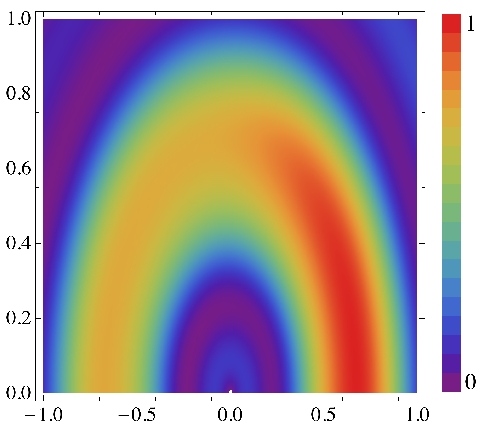}
\includegraphics[scale=0.4]{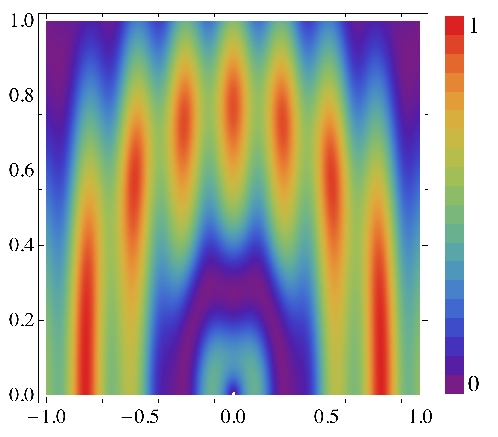}
\caption{\footnotesize{Normalized density plots of momentum distribution for detachment from $\textrm{H}^-$ ion with $\varphi=0$. Left panel -
distribution calculated with the use of $M^{(0)}+M^{(1)}$, right panel - standard SFA approach}}
\label{densplot1}
\end{figure}

\begin{figure}[here]
\centering
\includegraphics[scale=0.4]{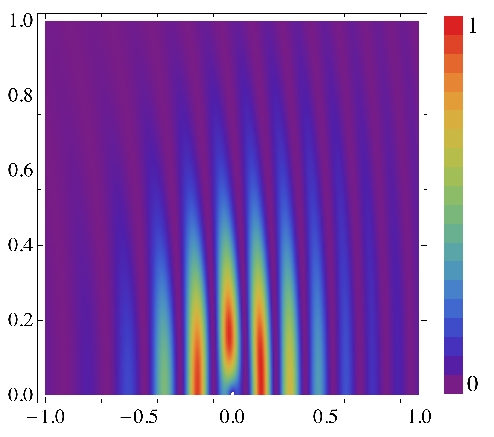}
\includegraphics[scale=0.4]{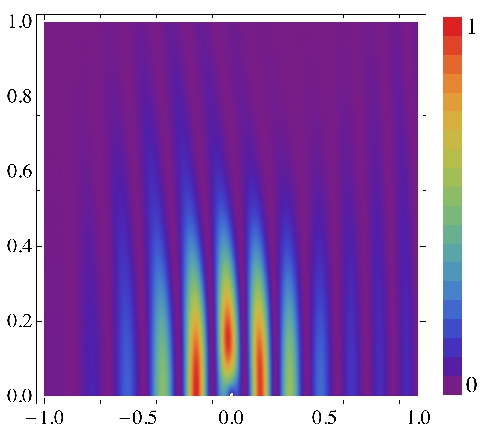}
\caption{\footnotesize{Normalized density plots of momentum distribution for detachment from $\textrm{H}^-$ ion with $\varphi=\pi/2$. Left panel -
distribution calculated with the use of $M^{(0)}+M^{(1)}$, right panel - standard SFA approach}}
\label{densplot2}
\end{figure}

Density plots of the probability distributions in the $(p_z,\,p_t)$ plane are shown in Fig. \ref{densplot1} for the CEP $\varphi=0$ and in Fig. \ref{densplot2} for $\varphi=\pi/2$. In the first case one notes substantial differences between the distribution obtained from formla \eqref{eq:17} and the standard SFA result, which could be also observed for this value of CEP in the $p_z$ distributions in the polarization direction. On the other hand, with CEP equal to $\pi/2$ results of the present calculations are very close to the standard SFA-approach, at least for the values of the pulse parameters used here.
\newpage
\section{Conclusions}
We have obtained manifestly gauge invariant expressions for the amplitudes of ionization by short and intense laser pulses in the dipole approximation. The derivation is based on the SFA-approach but, contrary to its standard formulation, all terms of the same order in atomic potential were consequently grouped together. This procedure allowed to eliminate gauge-dependent terms resulting in a new type of expansion with redefined nontributions of the "direct ionization" and "rescattering" - type ($M^{(0)}$ and $M^{(1)}$ respectively in \eqref{eq:17}). The approach proposed in this Letter is gauge-invariant order by order in the expansion for any gauge choice from the class of gauges defined by \eqref{eq:12}. This is in contrast with the approach used in ~\cite{Vanne} for the standard SFA approach, where for a class of gauges compatible with dipole approximation gauge-invariance can be achieved in the limit of infinite series. The present approach is based on two "natural" partitions of the total Hamiltonian: atomic Hamiltonian plus interaction with the laser field (Eq. \eqref{eq:1}], or Hamiltonian of an electron in the laser field plus atomic potential [Eq. \eqref{eq:3}]. These two partitions can be used for any gauge from the class \eqref{eq:12} with no necessity to use a notion of generalized field free Hamiltonian, as has been done in~\cite{Faisal3,Vanne}.

Numerical calculations of the detachment probabilities based on formulae \eqref{eq:17} show for zero CEP substantial differences in comparison with standard velocity gauge SFA, whereas for CEP equal to $\pi/2$ the differences are much smaller, at leat for the pulse parameters used in the present calculations. The issue of a relation between the standard SFA/rescattering approach, as well as the problem of convergence, require further examination. Preliminary estimations suggest that present approach is more suitable for very short pulses, basically in the subfemtosecond region.


\begin{thebibliography}{99}
    \bibitem{Keldysh} L.V. Keldysh, Zh. Eksp. Teor. Fiz. 47 (1964 ) 1945,\\L.V. Keldysh, Sov Phys-JETP {\bf 20} (1965) 137,
    \bibitem{Faisal} F.H.M. Faisal, J. Phys. B: At. Mol. Phys. 6 (1973) L89,
    \bibitem{Reiss} H.R. Reiss, Phys. Rev. A 22 (1980) 1786,
    \bibitem{Karnakov} B.M. Karnakov, V.D. Mur, S.V. Popruzhenko, V.S. Popov, Phys. Letters A 374 (2009) 386,
    \bibitem{Becker1} A. Becker, F.H.M. Faisal, J. Phys. B: At. Mol. Opt. Phys. 38 (2005) R1,
    \bibitem{Brabec} T. Brabec, F. Krausz, Rev. Mod. Phys. 72 (2000) 545,
    \bibitem{Milosevic} D.B. Milo\v{s}evi\'c, G.G. Paulus, D. Bauer, W. Becker, J. Phys. B: At. Mol. Opt. Phys. 39 (2006) R203,
    \bibitem{Milosevic1} D.B. Milo\v{s}evi\'c, W. Becker, M. Okunishi, G. Pr\"umper, K. Shimoda, K. Ueda, J. Phys. B: At. Mol. Opt. Phys. 43 (2010) 015401,
    \bibitem{Gazibegovic} A. Gazibegovi\'c-Busulad\v{z}i\'c, D.B. Milo\v{s}evi\'c, W. Becker, Phys. Rev. Letters 104 (2010) 103004,
    \bibitem{Faisal1} F.H.M. Faisal, G. Schlegel, J. Phys. B: At. Mol. Opt. Phys. 38 (2005) L223,
    \bibitem{Faisal2} F.H.M. Faisal, G. Schlegel, J. Mod. Optics 53 (2006) 207,
    \bibitem{BauerJ} J. Bauer, J. Phys. B: At. Mol. Opt. Phys. 41 (2008) 185003,
    \bibitem{BauerD} D. Bauer, D.B. Milo\v{s}evi\'c, W. Becker, Phys. Rev. A 72 (2005) 023415,
    \bibitem{Reiss1} H.R. Reiss, Phys. Rev. A 76 (2007) 033404,
    \bibitem{Bergues} B.Bergues, Yongfang Ni, H.Helm, J.Yu. Kiyan, Phys. Rev. Letters 95 (2005) 263002,
    \bibitem{Gazibegovic1} A.Gazibegovi\'c-Buzulad\v{z}i\'c, D.B. Milo\v{s}evi\'c, W.Becker, B.Bergues, H.Hultgren, I.Yu. Kiyan, Phys. Rev. Letters, 104 (2010) 103004,
    \bibitem{Kjeldsen} T.M. Kjeldsen, L.B. Madsen, Phys. Rev. A 71 (2005) 023411
    \bibitem{Cormier} E. Cormier, P. Lambropoulos, J. Phys. B: At. Mol. Opt. Phys. 29 (1996) 1667,
    \bibitem{Henneberger} W.C. Henneberger, Phys. Rev. Letters 21 (1968) 838,
    \bibitem{Madsen} L. B. Madsen, Phys. Rev. A 65 (2002) 053417,
    \bibitem{Lamb} W.E. Lamb, Phys. Rev. 85 (1952) 259,
    \bibitem{Starace} A.F. Starace, Phys. Rev. A 3 (1971) 1242,
    \bibitem{Fried} Z. Fried, Phys. Rev. A 8 (1973) 2835,
    \bibitem{Bassani} F. Bassani, J.J. Forney, A. Quattropani, Phys. Rev. Letters, 39 (1977) 1070,
    \bibitem{Kobe} D.H. Kobe, A.L. Smirl, Am. J. Phys. 46 (1978) 624,
    \bibitem{Kobe1} D.H. Kobe, Phys. Rev. A 19 (1979) 205,
    \bibitem{Rzazewski} K. Rz\c{a}\.zewski, R.W. Boyd, J. Mod. Optics, 51 (2004) 1137,
    \bibitem{Faisal3} F.H.M. Faisal, J. Phys. B: At. Mol. Opt. Phys. 40 (2007) F145,
    \bibitem{Vanne} Y.V. Vanne, A.Saenz, Phys. Rev. A 79 (2009) 023421,
    \bibitem{BeckerW} W. Becker, F. Grasbon, R. Kopold, D.B. Milo\v{s}evi\'c, G.G. Paulus, H. Walther, Adv. At. Mol. Opt. Phys. 48 (2002) 35,
    \bibitem{Tsakiris} G.D. Tsakiris, K. Eidmann, J. Meyer-ter-Vehn, F. Krausz, New Journal of Physics, 8 (2006) 19,
    \bibitem{Tarasevitch} A.P. Tarasevitch, R. Kohn, D. von der Linde, J. Phys. B: At. Mol. Opt. Phys. 42 (2009) 134006,
    \bibitem{Rohringer} N. Rohringer, R. London, Phys. Rev. A 80 (2009) 013809,
    \bibitem{Goulielmakis} E.Goulielmakis, M.Schultze, M.Uiberacker, M.Hofstetter, U.Kleineberg, F.Krausz, Acta Phys. Polon. A 112 (2007) 751.
\end{thebibliography}
\end{document}